\documentclass[14pt]{aastex631}
\def\lesssim{\mathrel{\hbox{\rlap{\hbox{\lower4pt\hbox{$\sim$}}}\hbox{$<$}}}}
\def\gtrsim{\mathrel{\hbox{\rlap{\hbox{\lower4pt\hbox{$\sim$}}}\hbox{$>$}}}}

\usepackage{graphicx}
\usepackage{float}

\def\d{\rho_{\rm 1}}

\def\so{\Sigma_{\rm 0}}
\def\B{\begin{equation}}
\def\E{\end{equation}}

\def\B{\begin{equation}}
\def\E{\end{equation}}
\def\O{\Omega}

\def\ni{\noindent}

\def\v{\vec}
\def\o{\omega}
\def\D{\Delta}

\def\v{\vec}

\def\ep{\epsilon}

\shorttitle{QPO and dissipation}
\shortauthors{Dezen et. al.}

\begin{document}

\title{Effects of Dissipation Physics on High-frequency Quasi-periodic Oscillations in Black Hole X-ray Binaries}

\author{Theodore Dezen}
\email{tdezen@sandiego.edu}
\affil{Department of Physics and Biophysics, University of San Diego, San Diego, CA 92110}
\author{Sergio Gomez}
\affil{Department of Physics and Astronomy, California State University, Los Angeles, CA 90032}
\author{Kathryn Anawalt}
\affil{Department of Physics and Biophysics, University of San Diego, San Diego, CA 92110}
\affil{Department of Electrical Engineering, University of San Diego, San Diego, CA 92110}


\begin{abstract}

\ni We numerically investigate the effects of black hole spin and local dissipation profiles on high-frequency quasi-periodic oscillation (HFQPO) observed in black hole X-ray binaries (BHXB). Our HFQPO power spectra arise from self-consistent calculations that do not rely on ad-hoc assumptions regarding disk geometry. Our models combine radiative transfer and disk vertical structure equations with input motivated by first-principles three-dimensional simulations. We found that HFQPO power spectra may be sensitive to spatial distribution of dissipation rates while the quality factors are more sensitive to black hole spin. We discuss the observational implications of our results in context of steep power law (SPL) spectra from BHXBs that are seen together with HFQPOs, and how QPO properties may be indicators of the underlying physical oscillations.

\end{abstract}

\section{Introduction}

\ni Milky Way black hole X-ray binaries (BHXBs) fall into several outburst states characterized by different spectral shapes and light curve variabilities \citep{mr06, dgk07}. In particular, BHXBs radiating at near or above the Eddington limit may display an energetically significant steep power law (SPL state, with photon index $\Gamma>2.4)$ tail that stretches from $\la 10 \ \rm keV$ all the way to hundreds of keV or even MeVs \citep{gr98, lw05}. BHBs in SPL state can simultaneously exhibit strong variability, termed high-frequency quasi-periodic oscillations (HFQPOs), in their hard X-ray light curves in $~10 - 30$ keV band. Detected HFQPOs \citep{ingram19} range from $~ 67$ Hz in GRS 1915+105 to hundreds of Hz in GRO J1655-40 \citep{bel12} and XTE J1550-564 \citep{miller01}. The physical origin of HFQPOs, especially their relationship to the SPL state, remain an open problem in astrophysics. 

\

\ni In a recent attempt to explain the SPL state in conjunction with HFQPOs, \cite{db14} observed that in a disk with non-zero magnetic torques at the innermost stable circular orbit (ISCO) \citep{ak00}, effective temperature would rapidly increase inward (decreasing distance) towards the black hole. The spectral peaks from different radii may then spread out sufficiently in photon frequencies to produce a SPL-like non-thermal tail in the full-disk spectrum. Such a model would naturally explain why HFQPO power spectra peaks become sharper (with higher quality factors $Q$) when restricted to integrating over higher and narrower energy bands since photons with energy $>10 \ \rm keV$ primarily come from a narrow radii range close to the black hole. However, more detailed radiative transfer calculations by \cite{df18} suggested that incorporating inner torque alone is a promising mechanism for HFQPO but may be insufficient to drive the SPL state. 

\

\ni On the other hand, earlier three-dimensional shearing box simulations with vertical stratification pointed to another potential explanation for the SPL state. In particular, calculations over a wide span of gas to radiation pressure ratios \citep{tur04, hir06, kro07, bla07, hkb09} and subsequent analysis \citep{bla11} indicated that the rate at which MRI turbulence  \citep{bh91, bh98} dissipates bulk kinetic energy peaks at approximately a pressure scale height away (in vertical direction) from the disk mid-plane. Later global simulations \citep{sc13, jia14} reinforce these local results and found corona-like structures above and below the disk. A hotter corona may Compton up-scatter photons that originated from the colder disk underneath to produce SPL-like spectra \citep{tb13}. More recently, \cite{huang23} also found higher temperature ($10^8-10^9$ K) corona structures in radiation MHD simulations that specifically modeled near-Eddington accretion disks around stellar mass black holes.

\

\ni Motivated by both local and global simulations, \cite{dem19} attempted to combine a hot corona  with the \cite{db14} model. Expanding upon previous studies that leveraged similar methodology \citep{dav05, bla06, dav09}, these authors connected simulations to observed spectra by incorporating simulation-motivated dissipation profiles into one-dimensional calculations that self-consistently couple disk vertical structure to full frequency-dependent radiative transfer. They concluded that increasing the fraction of accretion power dissipated in disk upper-layers while incorporating non-zero inner edge stresses may lead to models that simultaneously exhibit SPL and HFQPOs with quality factor $Q>2$.

\ 

\ni In this work, we systematically explore the effects of black hole spin and spatial distribution of dissipation on HFQPOs associated with the SPL state. Compared to previous efforts \citep{db14}, we significantly expand the parameter space by constructing disk models of near-Eddington accretion flows at multiple black hole spins and with several simulation-motivated dissipation profiles that correspond to dissipating different fractions of the accretion power near the photospheres. Our QPO power spectra is based on self-consistently coupling radiative transfer and vertical structure equations at each annuli to produce spectral models that fully account for effects of metal opacities and Comptonisation. In addition to shedding light on the dissipation process, the shearing box simulations referenced earlier also suggest that the $\alpha$-model \citep{ss73} relationship between pressure and total stress approximately hold \citep{bla11}. We therefore use $\alpha$-model equations with inner torque modifications described by \cite{ak00} to obtain the radial dependence of disk parameters that serve as input for our spectra and vertical structure calculations.

\

\ni We begin by outlining our calculation of synthetic QPO power spectra in Section $2$. In Section $3$, we describe our treatment of and assumptions regarding local dissipation rates, which are necessary for computing the QPO profiles. We present and discuss numerical results in Section $4$. Finally, we summarize and outline possible future work in Section $5$.

\section{Synthetic HFQPO Power Spectra}

\ni To construct a HFQPO power spectrum, we divide an accretion disk around a $7$ solar mass black hole into approximately $50$ annuli with the inner edge at ISCO. Following \cite{db14}, we assume that local variability amplitude is proportional to local X-ray flux, which is consistent with observations \citep{u01}. However, we caution that this assumption still awaits more detailed validation from radiative GRMHD simulations that can extract time and frequency-dependent light curves and hence variability as a function of distance from the black hole. We further assume each annulus contributes to the variability at the vertical epicyclic frequency $\O_z$ at that radius. At a particular radius, we then integrate the radiation flux within the desired energy range to obtain the HFQPO strength at the corresponding frequency, and scale by $r/r_g$ to account for scaling with emission area. We then normalize to total flux (from the entire disk) in the energy band and square to convert to power, resulting in dimensionless relative QPO power spectra $A(\O_z)$. We perform this process for disks accreting onto black holes with $a/M=0.4, 0.5, 0.6, 0.7, 0.8$ and $0.99$ and include $3$ different simulation motivated dissipation prescriptions (see next section) per spin.

\

\begin{figure}
\includegraphics[width=14cm, height=10cm]{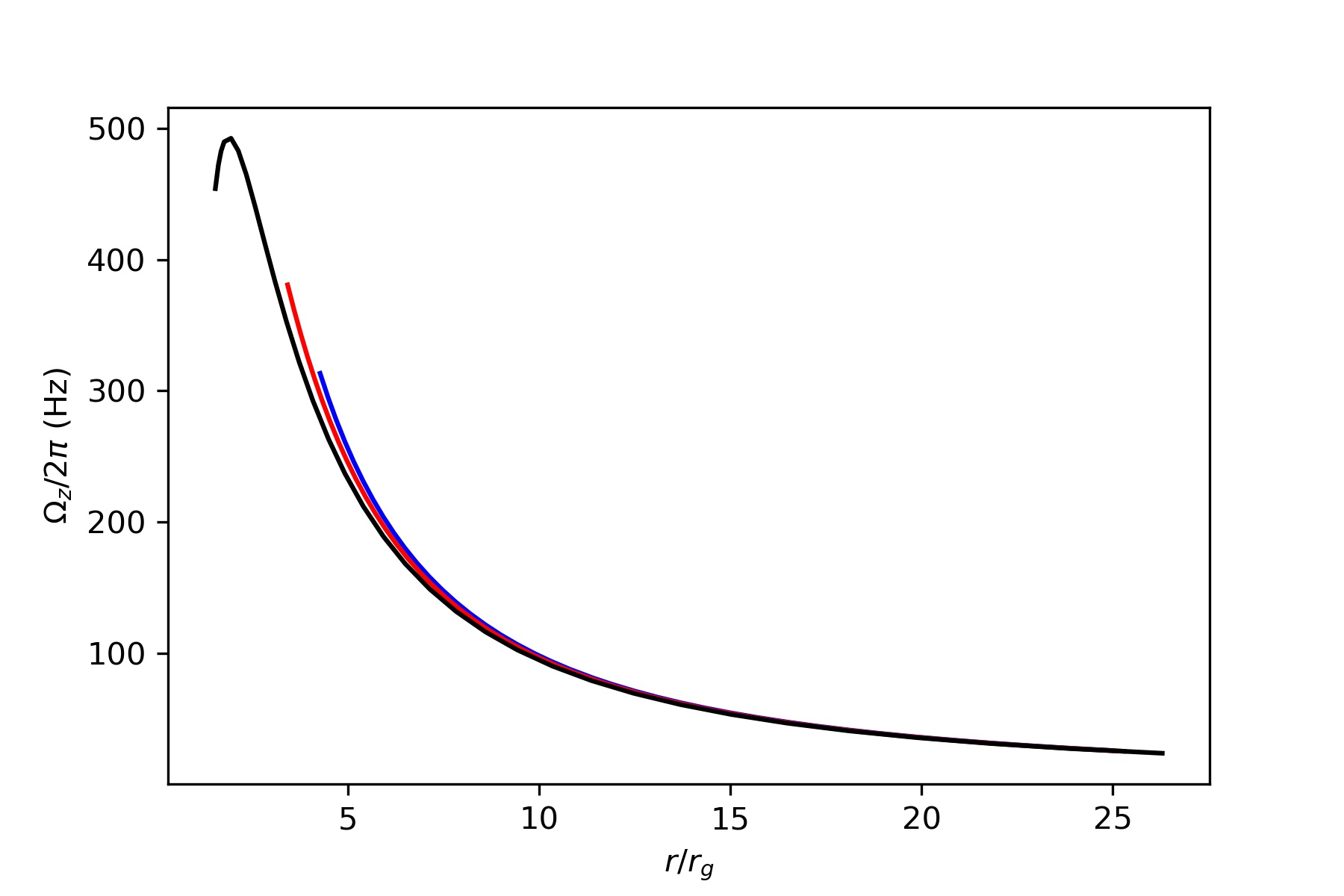}
\caption{Representative vertical epicyclic frequencies for black holes with spins $a=0.5$ (blue), $a=0.7$ (red) and $a=0.99$ (black). The inner end-point of each curve is the corresponding ISCO.} 
\label{epi}
\end{figure}

\ni Before proceeding, note that in Kerr spacetime the vertical epicyclic frequency $\O_z$ at coordinate distance $x=r/r_g$ from a black hole with spin $a$ is \citep{ts05}
\B
\O_z=\O_k\left(1-\frac{4a}{x^{3/2}}+\frac{3a^2}{x^2}\right)^{1/2},
\E
where 
\B
\O_k=\sqrt\frac{GM}{r_G^3}\frac{1}{x^{3/2}+a}
\E
is the Keplerian frequency. As illustrated in Figure \ref{epi}, $\O_z$ peaks at a radius outside of the ISCO at higher black hole spins and in such cases there are two radii that contribute to the same HFQPO frequency at above approximately $450$ Hz. We therefore may anticipate a stronger variability signal at these frequencies compared to spins where $\O_z$ is strictly monotonic. However, this enhancement would occur only if more detailed spectral calculations indicate $\O_z$ peaks at a distance $x=r/r_g$ reasonably close to the maximum in frequency integrated (over desired photon energy band) flux.

\

\ni To obtain the annuli fluxes necessary for constructing QPO profiles, we self-consistently compute the vertical structure and local emergent spectra using the one-dimensional stellar and accretion disk atmosphere code TLUSTY \citep{hl95, h98, h00}. TLUSTY treats a stationary plane-parallel system so that quantities such as density and temperature only depend on vertical distance from the mid-plane. The code solves the relativistic equations governing disk structure, statistical equilibrium of level populations along with that of full multi-frequency and angle dependent radiative transfer. Earlier TLUSTY-based spectral fitting produced promising but mixed results in both AGN \citep{bla01} and BHB \citep{mc06} but later studies that leveraged TLUSTY's input flexibility to incorporate newer physics insights made strides in understanding various aspect of BHXB accretion such as the SPL state \citep{tb13, dem19}, slim disk models \citep{sdm13} spectral hardening \citep{dav19} and emission from plunging region \citep{zhu12}. Interested readers can consult several other previous TLUSTY-based spectral and disk structure studies \citep{dav05, dav09, df18} for more details regarding the code and its applications.

\

\begin{figure}
\includegraphics[width=9cm, height=6cm]{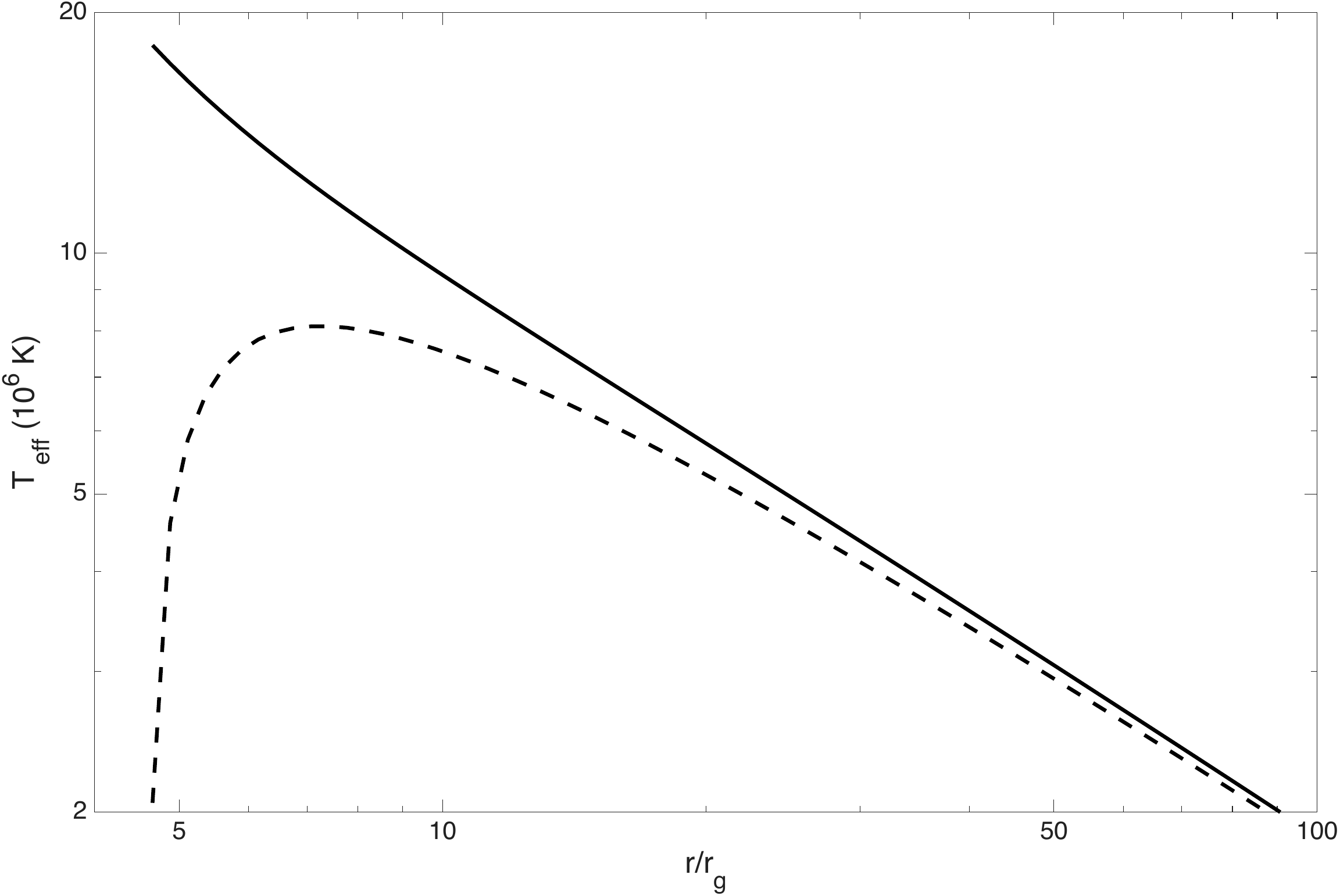}
\includegraphics[width=9cm, height=6cm]{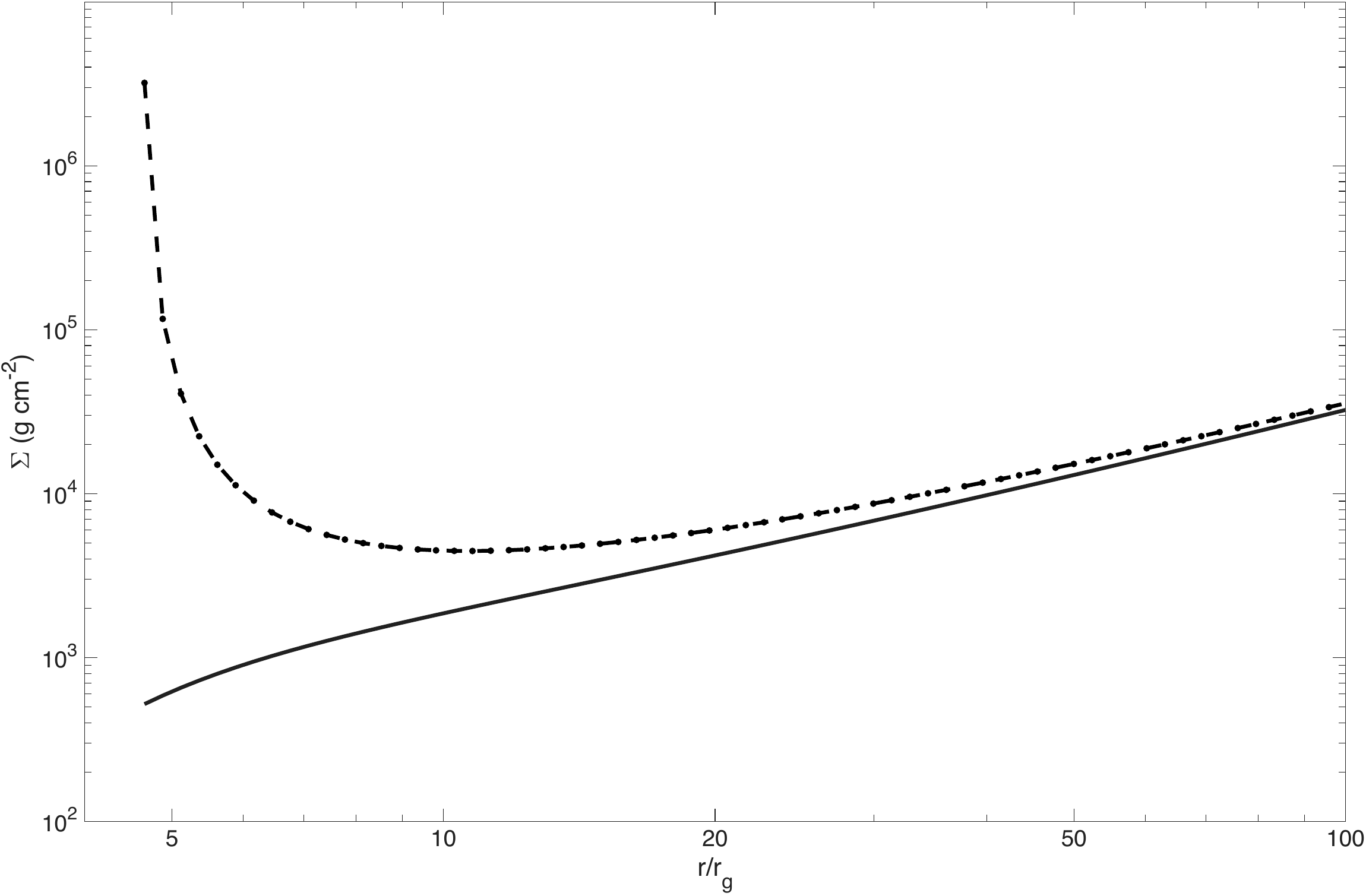}
\caption{Representative radial disk profiles as function of distance from black hole normalized to gravitational radii. Left and right panels are effective temperature $T_{\rm eff}$ and total surface density $\Sigma$, respectively. Solid and dotted curves represent calculations with non-zero inner torque and without, respectively.}
\label{ak}
\end{figure}

\ni At each annulus, TLUSTY requires the effective temperature $T_{\rm eff}$, total vertical column density $\Sigma_0$ and Keplerian frequency $\O_k$ as inputs, all of which depend on distance $x\equiv r/r_g$ from the black hole, where $r_g$ is the gravitational radius. We compute these parameters incorporating effects of magnetic stresses at the inner disk edge with an efficiency enhancement factor $\Delta\ep$ \citep{ak00, db14} so that the normalized accretion rate is
\B
\dot{m}=\frac{\ep\dot{M}c^2}{L_{\rm Edd}}=\frac{(\ep_0+\Delta\ep)\dot{M}c^2}{L_{\rm Edd}},
\E 
where $M$ is the black hole mass and $\ep_0$ is the radiative efficiency with the standard zero inner torque inner boundary condition. For comparison with previous work \citep{df18, dem19}, we choose $\Delta\ep=0.1$, which correspond to luminosity $L\approx L_{\rm edd}$. As illustrated in Figure \ref{ak}, unlike standard thin disks \citep{ss73, nt73}, our effective temperature $T_{\rm eff}$ rises sharply as $r/r_g$ decreases rather than drop to $0$, which increases the photon energy separation of annuli spectral peaks per unit radial separation and hence such models can potentially become natural filters for HFQPOs \citep{db14, dem19}. 

\ni Following \cite{ak00}, we also take viscosity parameter $\alpha$ \citep{ss73} to be spatially uniform, but some simulations including \cite{huang23} found effective $\alpha$ to vary with radius. However, GRMHD simulations such as \cite{noble09} produced radial flux profiles that agreed well with \cite{ak00} models that  our effective temperature and surface density profiles are also based on. More recent GRMHD simulations such as \cite{zhu12} that included region within the ISCO produced radial disk profiles (in quantities such as $T_{\rm eff}$ and $\Sigma$) that agreed with the key qualitative aspects of our work - namely that unlike \cite{nt73} disks, the surface density $\Sigma$ does not sharply increase and effective temperature $T_{\rm eff}$ does not plunge towards the black hole.  Finally, more recent GRMHD calculations \citep{zhang25} at higher resolution also broadly support these findings even though their radial disk profiles differ in quantitative details from ours.

\section{Dissipation Profiles}

\ni In addition to annuli parameters such as effective temperature, surface density and orbital frequency, TLUSTY also needs height-dependent (or distance from mid-plane) local dissipation rates that quantify how the gravitational potential energy lost by the accreting material is converted to thermal energy that ultimately heats the plasma. We described the electron-ion fluid with a single gas temperature $T_{\rm g}$ since the disk scale height is much greater than collision mean free path between particle species. However, gas and radiation may not be in thermal equilibrium. We further consider only vertical radiative diffusion in energy transport and hence relate the dissipation rate per unit volume $Q$ to the total frequency-integrated radiative flux at vertical distance $z$ away from the mid-plane via
\B
Q(z)=\frac{dF}{dz}.
\label{qdfdz}
\E
We do not explicitly include magnetic fields in this exploratory study, however our dissipation profiles are motivated by MHD simulations and hence reflect the effects of magnetic pressure on disk structure. Moreover, \citep{dav09} found that the expected hardening of emergent spectra due to magnetic support against vertical gravity is largely negated by the softening effects of density inhomogeneities in disk upper-layers.

\

\begin{figure}
\includegraphics[width=12cm, height=8cm]{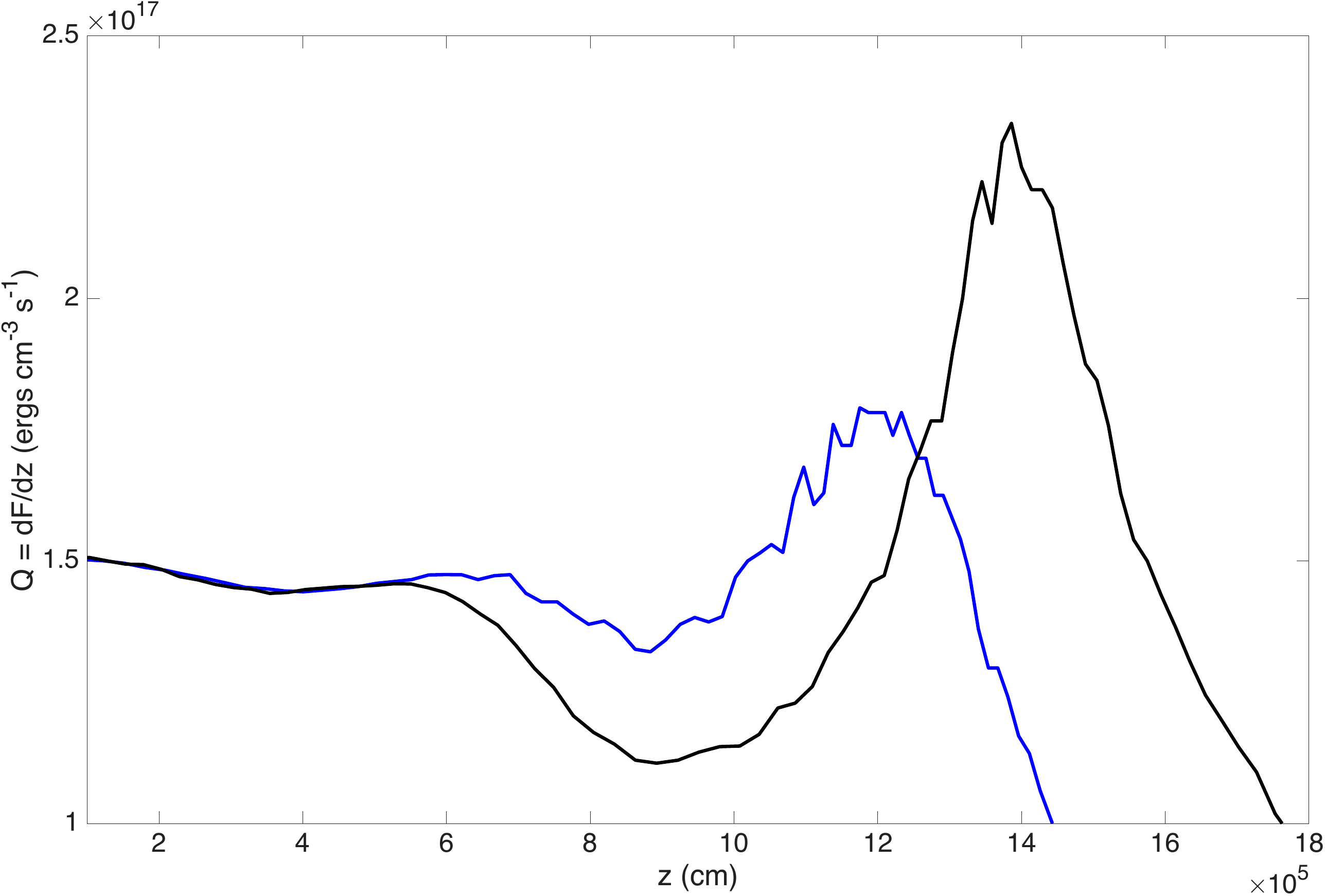}
\caption{Sample dissipation rates as functions of height $z$ (measured with mid-plane at $z\equiv 0$) for an annulus at $r/r_g\approx 4$ in a disk accreting onto an $a/M=0.7$ black hole. The blue and black curves are based on $\zeta=0.1$ and $\zeta=0.03$, respectively. The gas pressure scale heights here is approximately $1.1\times10^6$ cm and $1.3\times 10^6$ cm, for models represented by blue and black curves, respectively.}
\label{qz}
\end{figure}

\ni To study the effects of dissipating different fractions of the accretion power into the spectral forming regions of disk upper-layers, we adopt as input for TLUSTY several dissipation profiles that are power-laws when expressed in terms of fractional surface density $\Sigma/\so$ so that
\B
\frac{Q\Sigma}{\rho}=-\Sigma\frac{dF}{d\Sigma}=F_0\zeta\left(\frac{\Sigma}{\so}\right)^{\zeta},
\label{dis2}
\E
where $\rho$ is the mass density. The column density $\Sigma$ at vertical distance $z$ ($z=0$ is midplane) from the mid-plane is defined such that $\Sigma(z\rightarrow\infty)\rightarrow0$ and $\Sigma_0=\Sigma(z=0)$. To investigate the interplay between SPL spectra and HFQPO, we chose $\zeta=0.1$ because it is the highest value that gave rise to energetically significant non-thermal spectral tails \citep{tb13,dem19}. We also construct models with $\zeta=0.03$, below which TLUSTY would not capture all of the expected radiative flux as convergence considerations limited the minimum surface density of the computational domain. For comparison, we also perform all calculations for $\zeta=0.2$ as this is the best fit to the time and horizontally averaged dissipation per unit mass measured from the local simulations of \cite{hkb09}. Setting $\zeta$ to $0.03$ or $0.1$ in equation (\ref{dis2}) results in significantly higher dissipation at the disk surface layers than using $\zeta=0.2$. As illustrated in Figure (\ref{qz}), these profiles have peaks slightly above a gas pressure scale height (approximately $1.1$ and $1.3\times10^6 \ {\rm cm}$ for blue and black curves, respectively) from the disk mid-plane as a function of $z$, consistent with findings from global simulations such as \cite{jia14} and \cite{jia16}. More recently, \cite{huang23} found azimuthal and time-averaged dissipation profiles that have peaks away from disk mid-plane in their lower accretion rate $\dot M\approx 0.01\dot M_{\rm Edd}$ calculation. On the other hand, dissipation in their higher accretion rate $\dot M\approx 0.8\dot M_{\rm Edd}$ and $0.9\dot M_{\rm Edd}$ runs do not show clear peaks but are nonetheless still extended above and below the mid-plane, suggesting that there may be significant variation of dissipation profiles as function of accretion rate that warrants future investigations. Finally, Monte Carlo spectral calculations \citep{mills24} based on simulations that included slightly above Eddington runs, which had similar initial conditions and resulting disk structure as \cite{huang23}, further suggested that our vertical dissipation prescriptions can lead to energetically significant power-law spectral tails.

\

\ni Before proceeding, note that we use thin disk radial quantities as inputs to TLUSTY. On the other hand, radiation pressure dominated systems, especially those at near or higher than Eddington limit may be better described by slim disks \citep{ab88} where both radial and vertical energy transport are important. More specifically, \cite{sa11} found that the resulting radial dependence of quantities such as $\Sigma_0$ and $T_{\rm eff}$ may differ noticeably compared to thin disks. However, \cite{sdm13} generated TLUSTY models that approximately agreed with slim disk based radial quantities and found that the two approaches only differed slightly in spectra even for near Eddington flows. We therefore believe that thin disk models that incorporate simulation motivated dissipation profiles are sufficient for this initial numerical exploration of QPO power spectra.

\section{Results and Discussion}

\begin{figure}
\includegraphics[width=9cm, height=6cm]{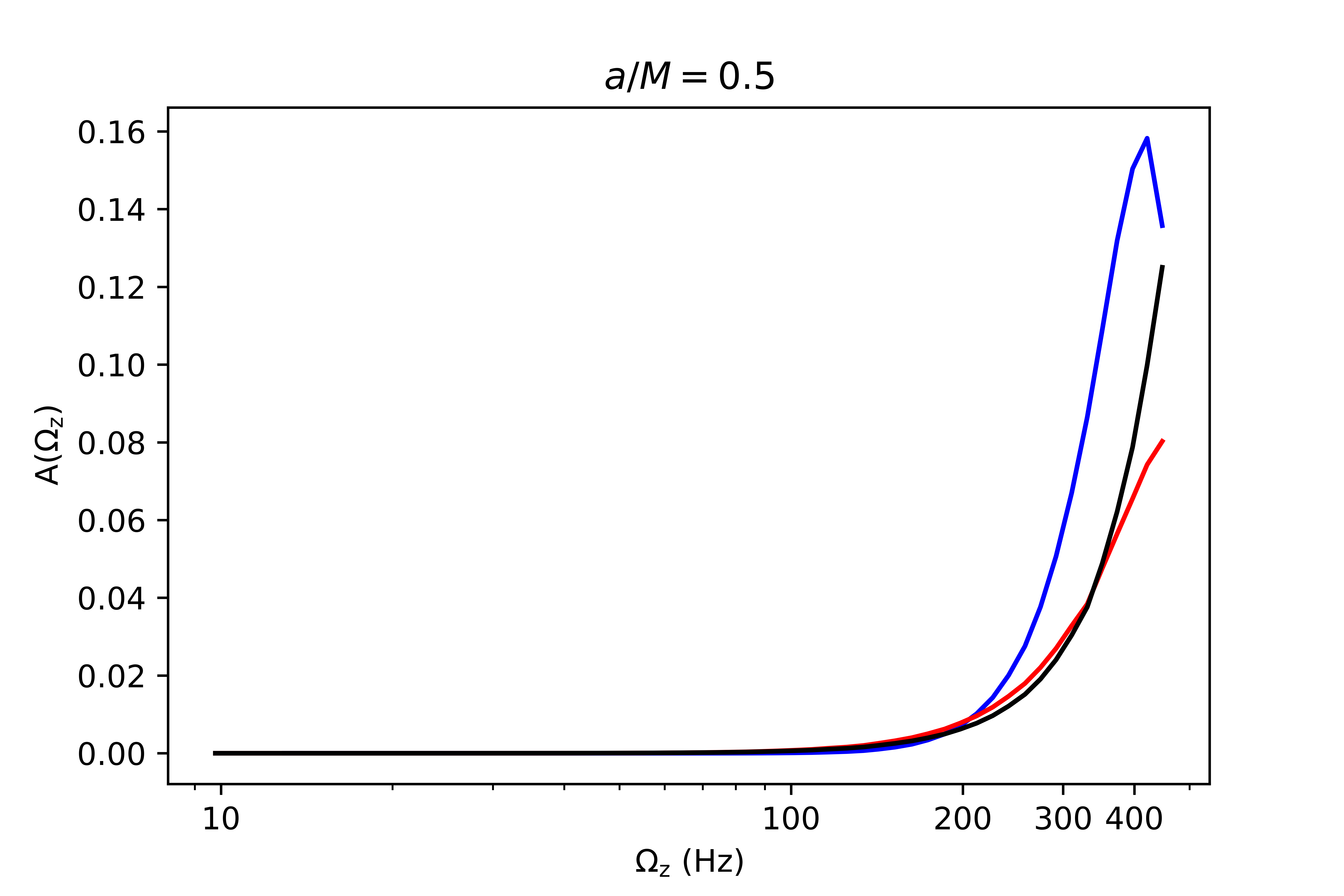}
\includegraphics[width=9cm, height=6cm]{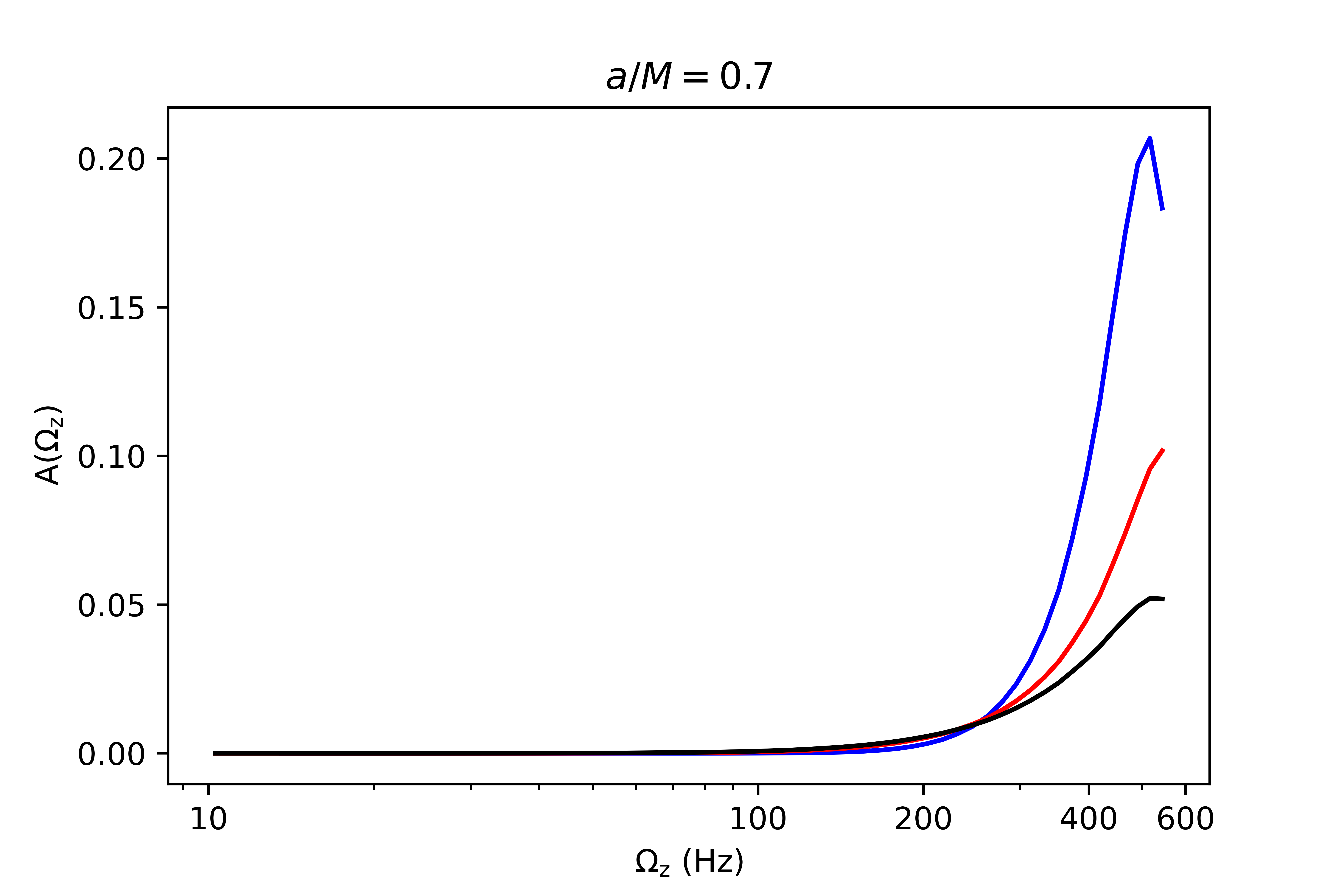}
\includegraphics[width=9cm, height=6cm]{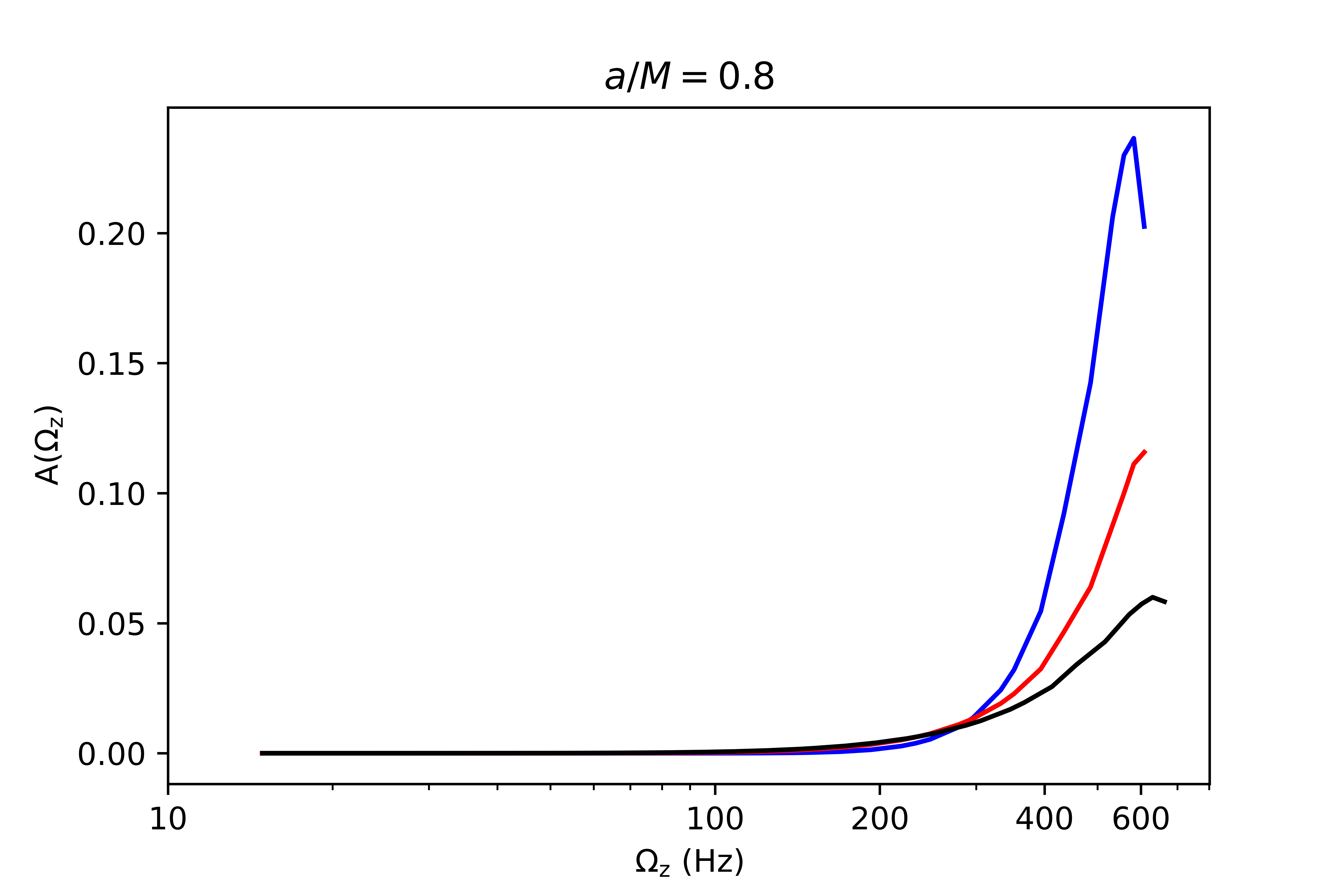}
\includegraphics[width=9cm, height=6cm]{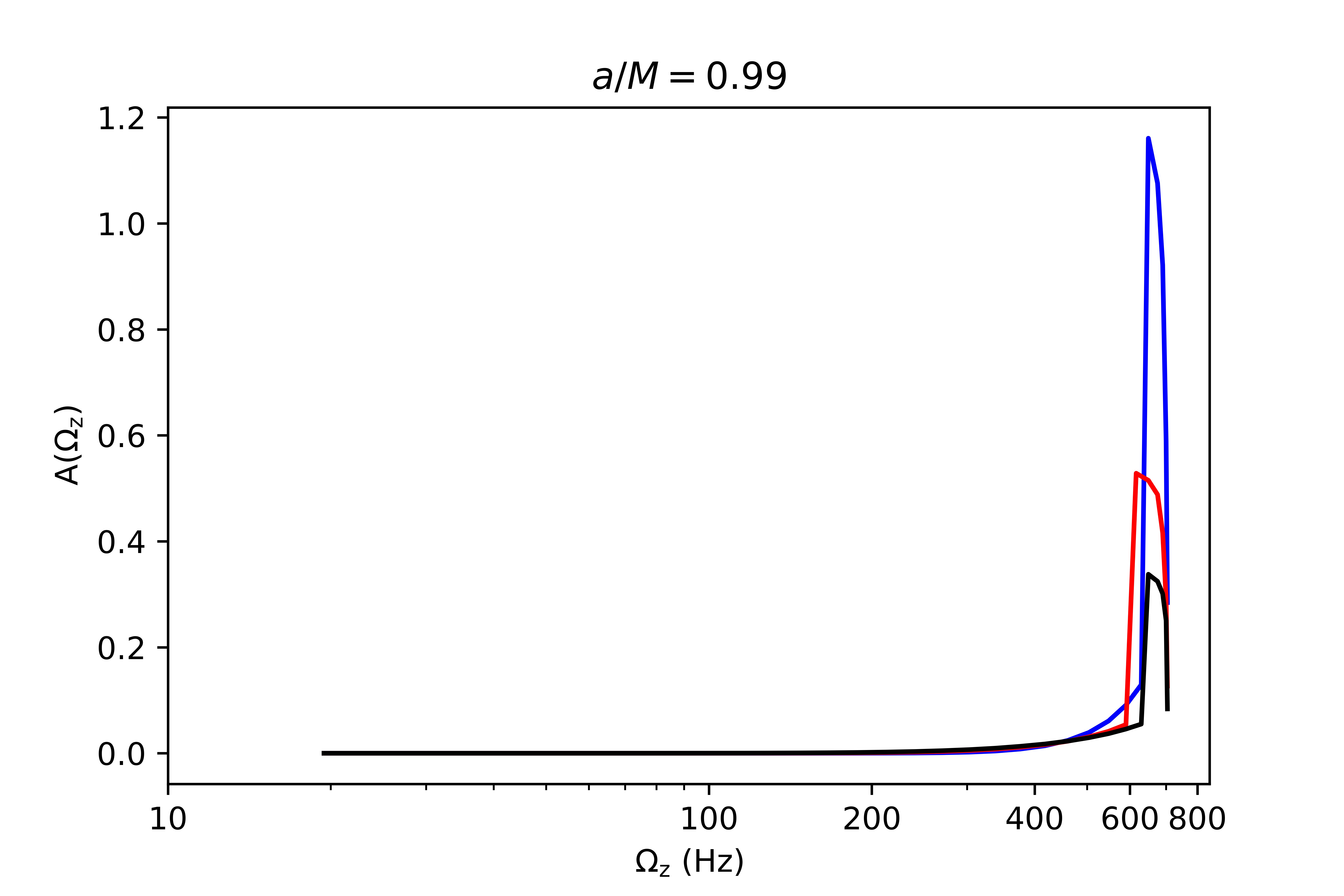}
\caption{Sample synthetic QPO power spectra for accretion disks around $a/M=0.5$ (top left), $0.7$ (top right), $0.8$ (bottom left) and $0.99$ (bottom right) black holes. Blue, red and black curves denote $\zeta=0.2, 0.1$ and $0.03$, respectively. All calculations here integrated over $10$ - $30$ keV in photon energies.}
\label{localqpo}
\end{figure}

\begin{figure}
\includegraphics[width=9cm, height=6cm]{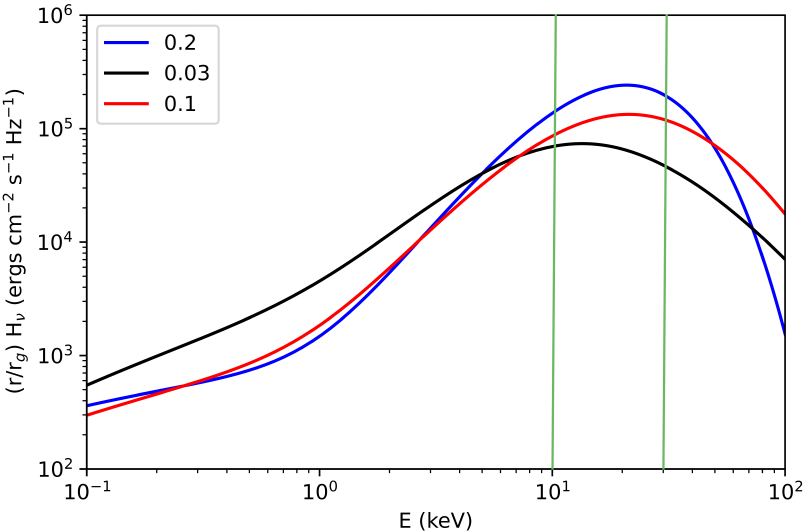}
\includegraphics[width=9cm, height=6cm]{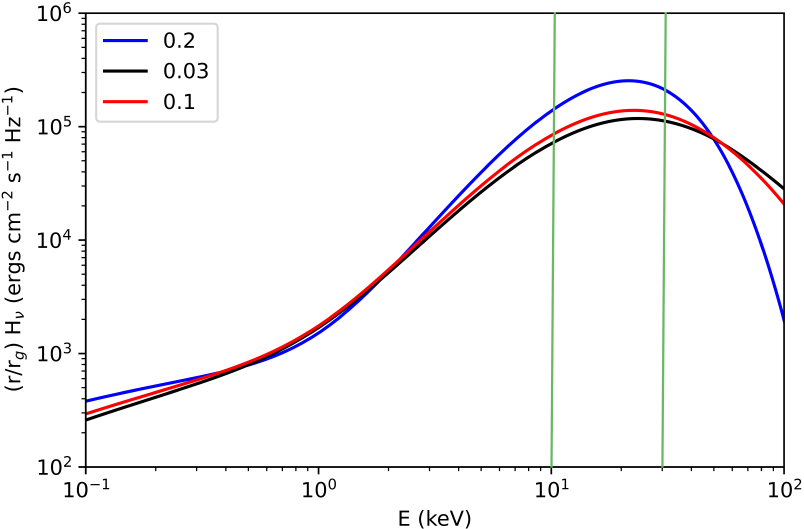}
\includegraphics[width=9cm, height=6cm]{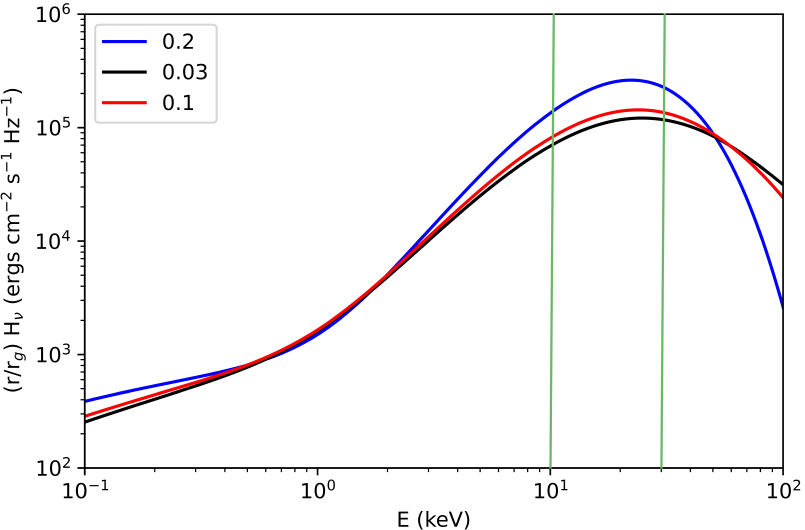}
\includegraphics[width=9cm, height=6cm]{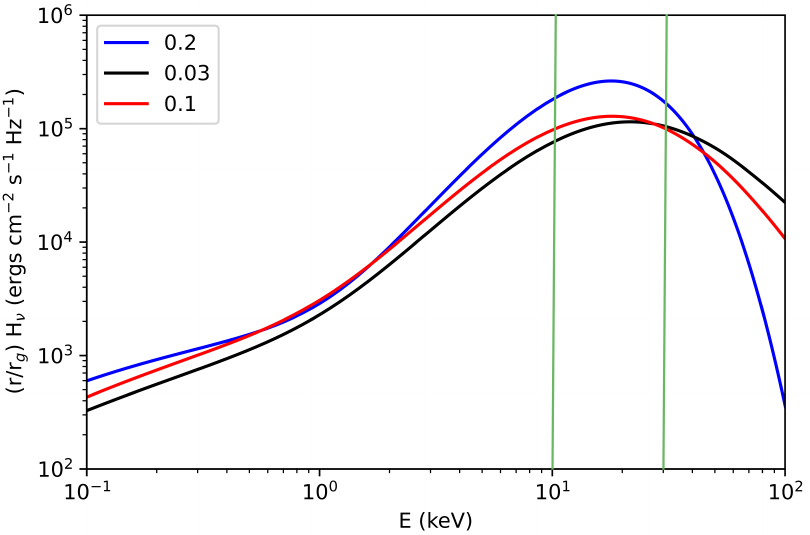}
\caption{Sample emission area weighted photon spectra from annuli that correspond to vertical epicyclic frequencies at the peak of QPO power spectra. The panels represent accretion disks around $a/M=0.5$ (top left), $0.7$ (top right), $0.8$ (bottom left) and $0.99$ (bottom right) black holes. Blue, red and black curves denote $\zeta=0.2, 0.1$ and $0.03$, respectively. Finally, vertically green lines indicate photon energies of $10$ keV and $30$ keV, respectively.}
\label{localspec}
\end{figure}

\begin{figure}
\includegraphics[width=15cm, height=10cm]{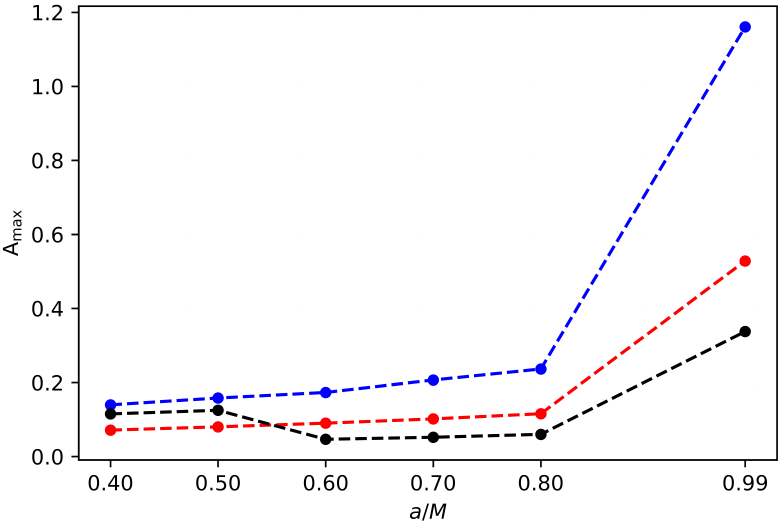}
\caption{Spin dependence of maximum relative QPO amplitude for accretion onto black holes. Blue, red and black denote dissipation prescriptions with $\zeta=0.2, \ 0.1$ and $0.03$, respectively.}
\label{peak}
\end{figure}

\ni We first present sample relative synthetic QPO power spectra $A(\O_z)$ from accretion disks around black holes with $a/M=0.5, 0.7, 0.8$ and $0.99$. Figure (\ref{localqpo}) indicates that both QPO width (and hence quality factor $Q$) and strength vary with local dissipation profile. As summarized in Figure (\ref{peak}), for all spins besides $a/M=0.4$ and $0.5$, QPO strength decreases with increasing accretion power (or decreasing $\zeta$ in equation \ref{dis2}) dissipated in disk upper layers. This result appears counterintuitive as it is reasonable to expect adding energy to the region near annuli photospheres would generate harder spectra with more emitted power $10 - 30$ keV range. However, Figure (\ref{localspec}) indicates that dissipating a larger fraction of accretion power in disk upper layers resulted in even harder annuli spectra where a significant number of photons up scattered into energetically significant power-law tails, leaving less flux between $10-30$ keV compared to annuli with $\zeta=0.2$, especially closer to the black hole. We further visualize these results in Figure (\ref{peak}), which shows that maximum relative QPO amplitude $A_{\rm max}$ tends to increase with black hole spin regardless of dissipation prescription. 

\

\begin{figure}
\includegraphics[width=6cm, height=4cm]{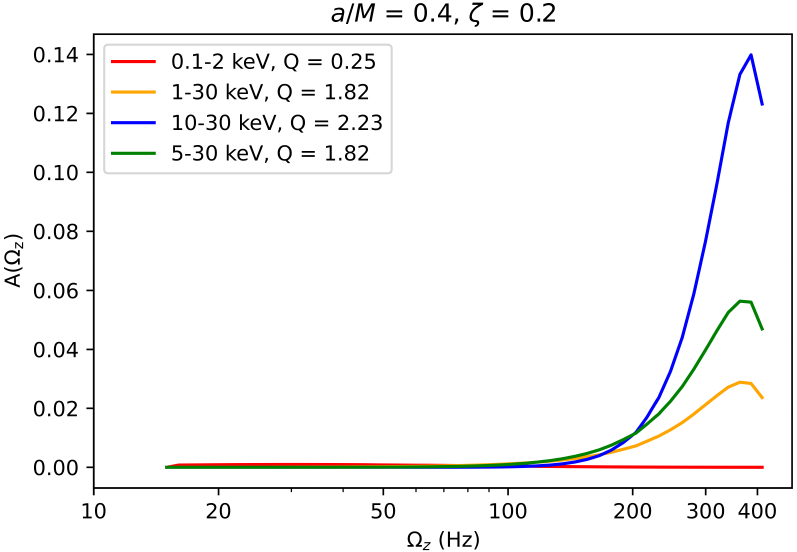}
\includegraphics[width=6cm, height=4cm]{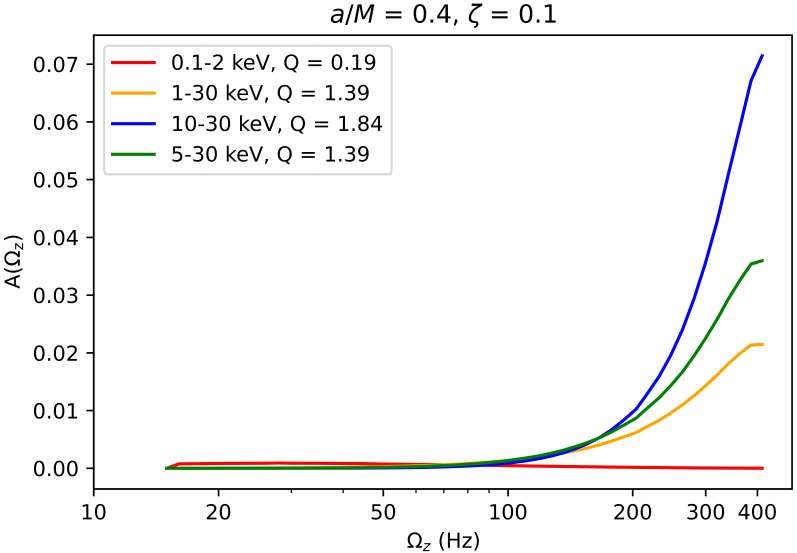}
\includegraphics[width=6cm, height=4cm]{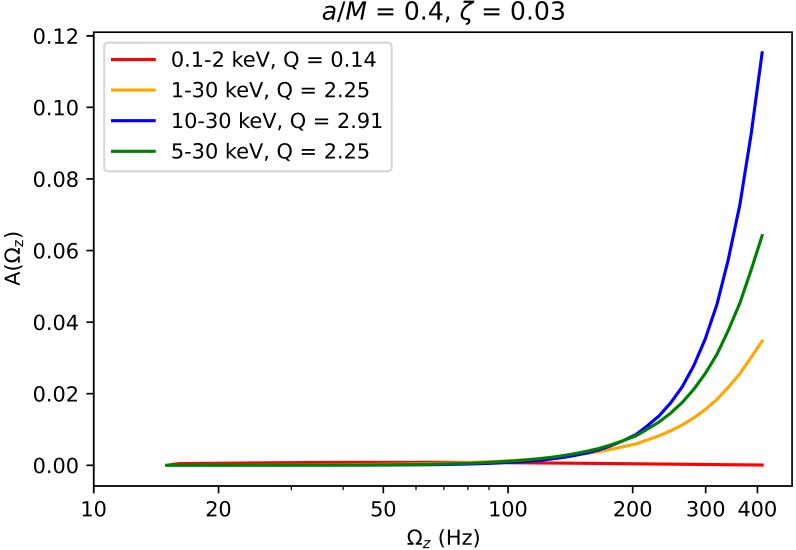}
\includegraphics[width=6cm, height=4cm]{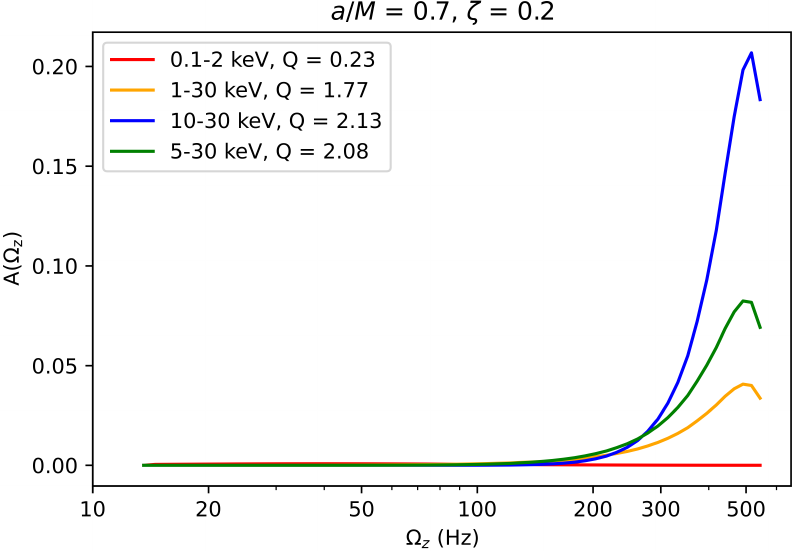}
\includegraphics[width=6cm, height=4cm]{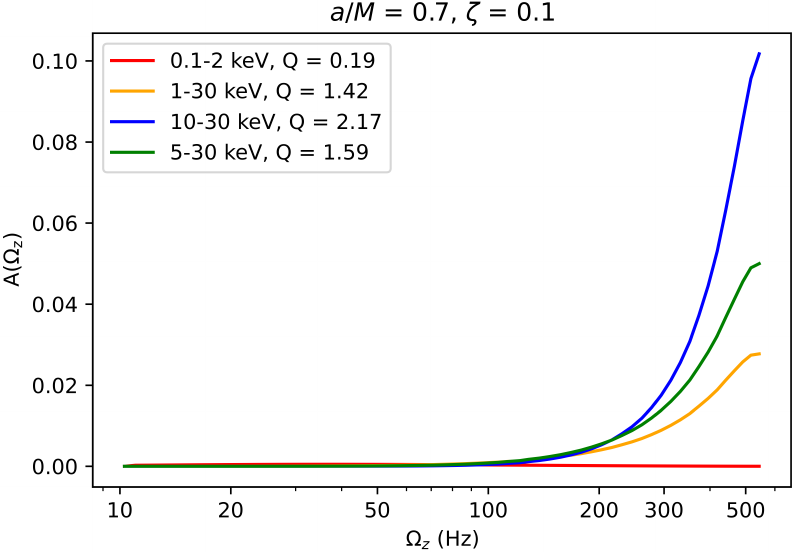}
\includegraphics[width=6cm, height=4cm]{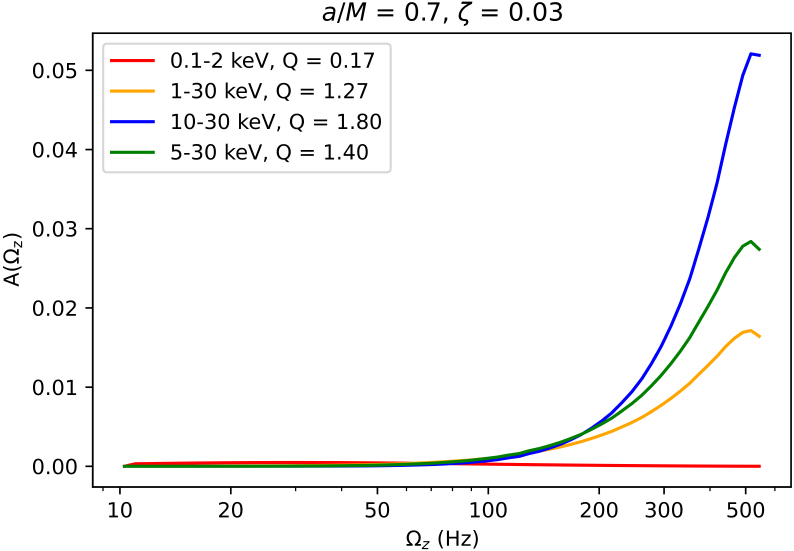}
\includegraphics[width=6cm, height=4cm]{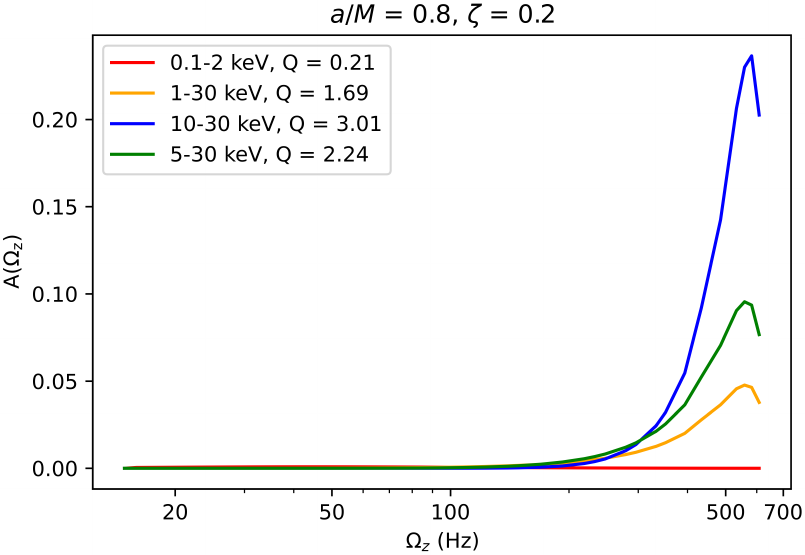}
\includegraphics[width=6cm, height=4cm]{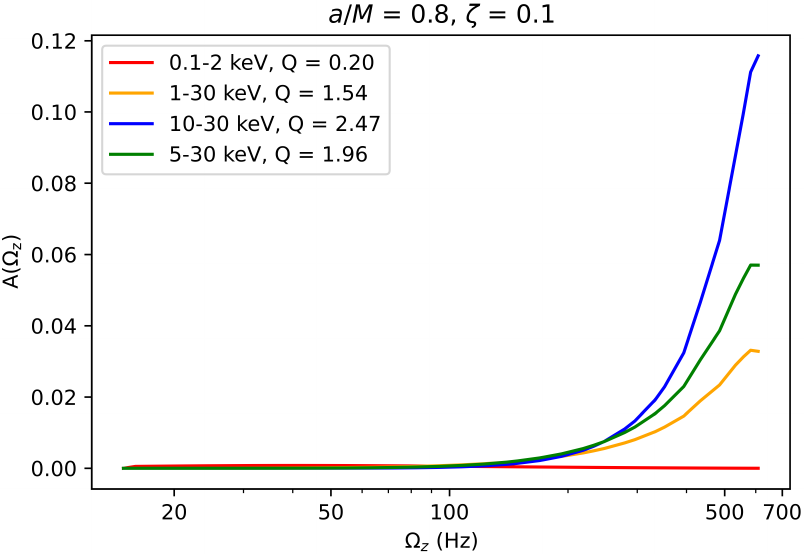}
\includegraphics[width=6cm, height=4cm]{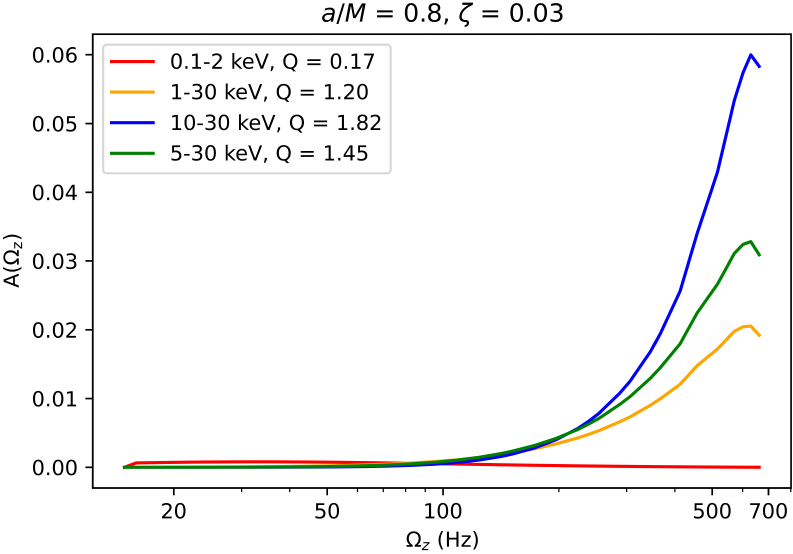}
\includegraphics[width=6cm, height=4cm]{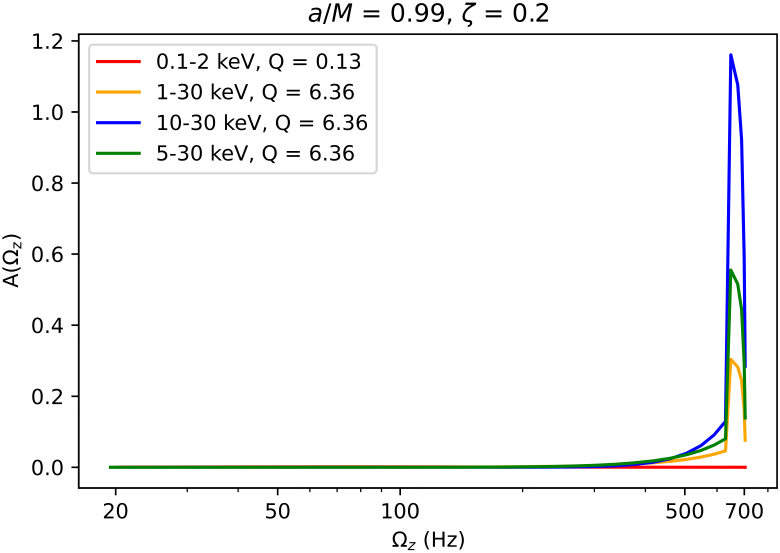}
\includegraphics[width=6cm, height=4cm]{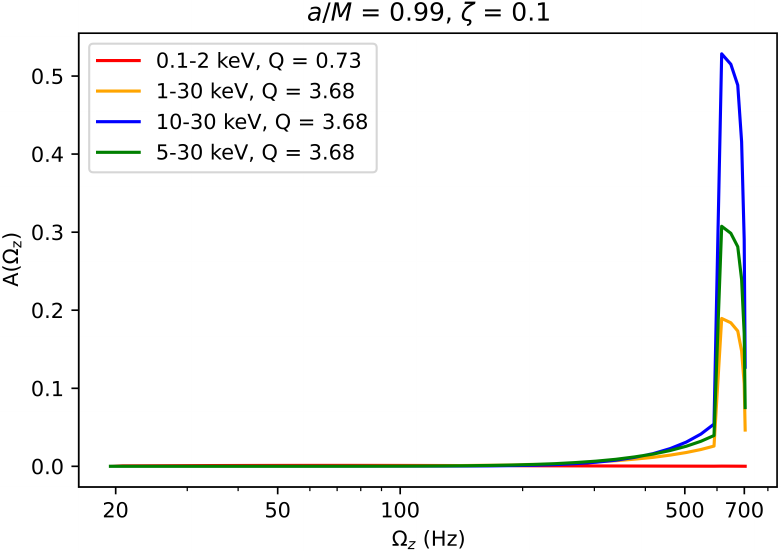}
\includegraphics[width=6cm, height=4cm]{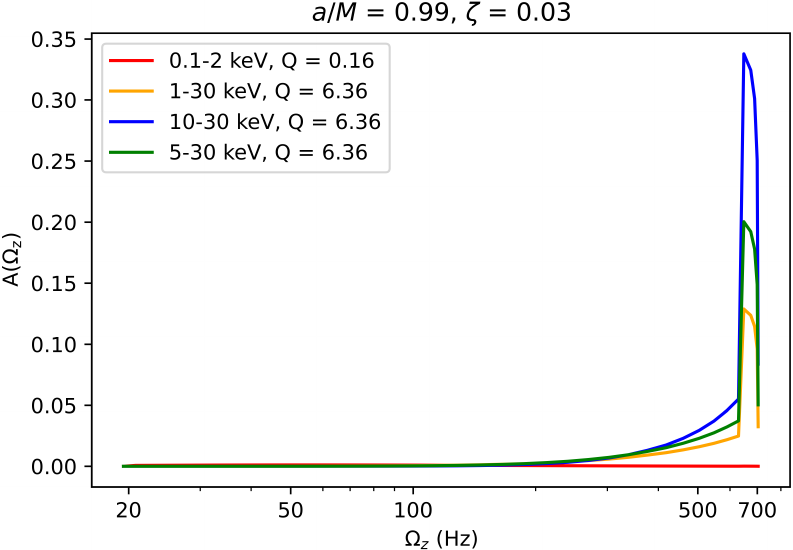}
\caption{Sample relative QPO power spectra from integrating over different photon energy bands.}
\label{photonrange}
\end{figure}

\ni \cite{db14} argued that disk models with non-zero magnetic torques at inner edge should exhibit broader QPO signals with lower quality factors $Q$ as we integrate over wider and lower photon energy bands, a conclusion supported at least qualitatively by \cite{dem19}. Our calculations broadly agree with these previous studies that QPO strength is highest for $10-30$ keV band in all cases as illustrated in Figure (\ref{photonrange}). On the other hand, we found that QPO quality factors do not depend strongly on photon energy range, except that no appreciable QPO signal exists in $0.1 - 2$ keV band, which is unsurprising. The dependence on dissipation profiles is noticeable but not drastic. However, for all spins other than $a/M=0.99$, models with $\zeta=0.03$ consistently gave $Q<2$. 

\

\ni We also found that accretion disks around black holes with higher spin should exhibit both stronger (higher $A_{\rm max}$) and sharper (higher $Q$) HFQPOs with higher quality factors, which in our model is primarily because the vertical epicyclic frequency $\O_z$ peaks outside of ISCO near maximally spinning black holes. It is hence possible that photons from multiple radii contribute to QPO power spectrum at the same variability frequency. Specifically, we obtained $Q\approx 1 - 3$ for all calculations except those at $a/M=0.99$, where $Q\approx 4 - 6$ regardless of spatial distribution of local dissipation rate in these disks. 

\

\begin{figure}
\includegraphics[width=12cm, height=8cm]{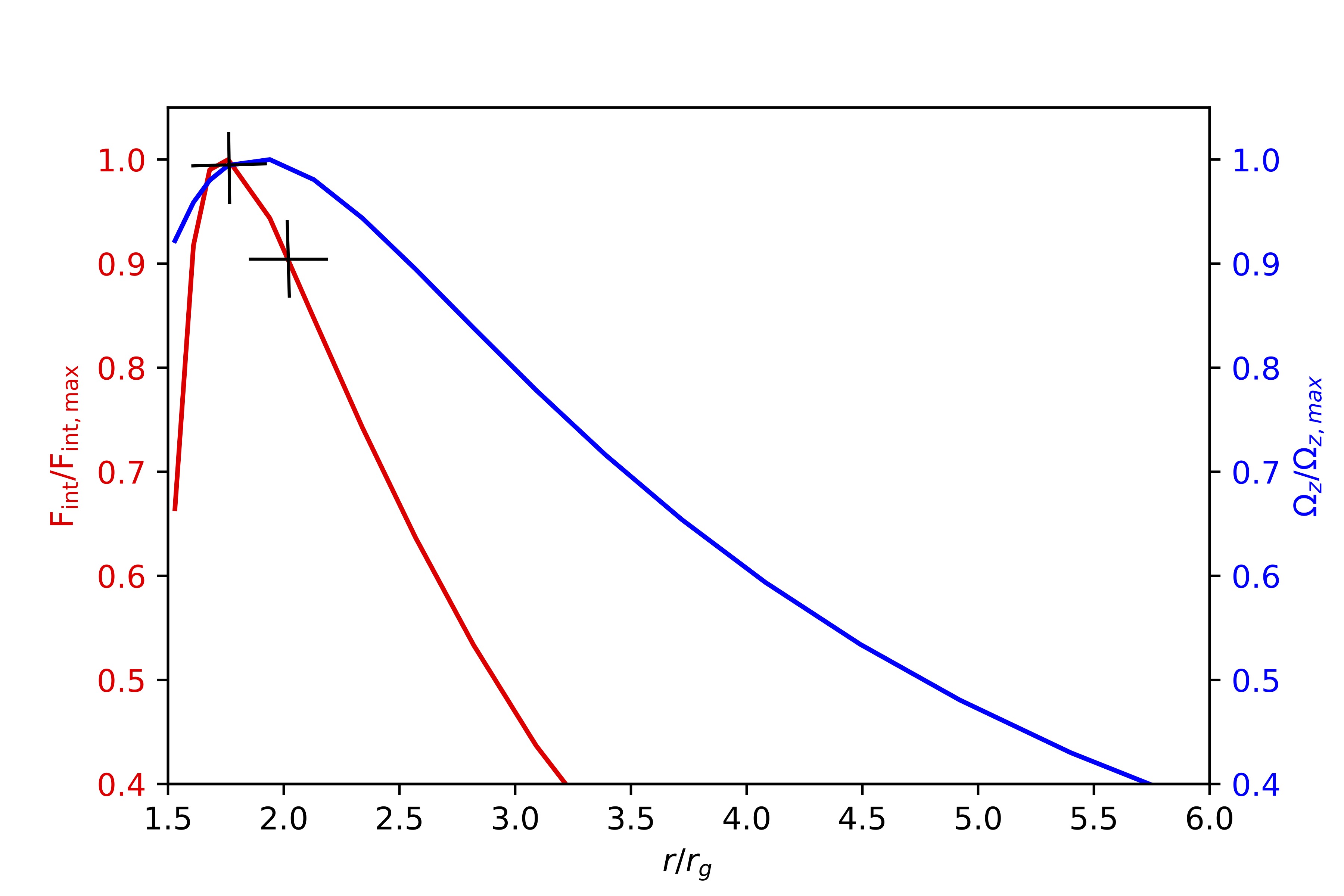}
\caption{Frequency integrated flux in $10-30$ keV band (red) and vertical epicyclic frequency (blue) as functions of distance from black hole $r/r_g$ for accretion onto a maximally spinning black hole with dissipation profile $\zeta=0.2$. Both quantities are normalized to their maximum values.}
\label{double}
\end{figure}

\ni More generally, we found that the QPO profiles computed from local annuli spectra increased in quality factor $Q$ with black hole spin regardless of dissipation profile. Specifically, our models consistently achieved $Q>3$ over a several photon energy bands with maximally spinning black holes. This is again mainly a result of our underlying model that HFQPOs are manifestations of vertical epicyclic modes, whose frequencies peak slightly outside of ISCO for nearly maximally spinning black holes. We further illustrate this process in Figure (\ref{double}) using the $a/M=0.99$ and $\zeta=0.2$ model, where the black crosses show an example in which fluxes from two different radii contribute to QPO power at the same epicyclic, and hence variability, frequency.

\

\ni Our work may thus point to a further observational test on underlying HFQPO physical models provided other methods constrain black hole spin. Finding BHXBs containing near maximally spinning black holes that also exhibit sharp HFQPOs with high quality factor ($Q\gtrapprox4$) would suggest that the vertical epicyclic mode may be the underlying fluid oscillation. Conversely, if observations can otherwise independently limit oscillation modes to those whose frequencies peak outside of ISCO, it may be possible to further constrain black hole spin with high quality factor HFQPOs. Moreover, our $a/M=0.99$ models are the only ones to exhibit quality factors higher than $5 - 6$, consistent with observations from, for example, GRO-$1655-40$ \citep{mr06}. 

\section{Conclusion and Future Work}

\ni We systematically explored the dependence of HFQPO power spectra on local dissipation prescription and black hole spin. Following and expanding upon previous radiative transfer calculations \citep{dav05,tb13,df18,dem19}, we computed QPO power spectra using disk models that self-consistently solved radiative transfer and vertical structure equations. In turn, these disk models leverage local dissipation physics motivated by first-principles three-dimensional local and global simulations. 

\

\ni While our quality factors depend most strongly on black hole spin, we found HFQPO amplitude to grow significantly with increasing fractions of accretion power dissipated near disk photospheres. It is therefore possible that only a certain range of local dissipation in disk upper-layers can produce HFQPO amplitudes that agree with observations. However, a more detailed study that fully incorporates relativistic transfer effects as seen by distant observers is necessary to gain further quantitative conclusions. Building on previous relativistic transfer algorithms by \cite{agol97}, an effort to this effect is underway.

\

\ni Another promising area for future work is to break away from modified thin-disk radial solutions \citep{ak00} because near or above Eddington, radiation pressure dominated accretion disks may be geometrically thick and hence better described by slim disk models \citep{ab88}. \cite{sa09a} and \cite{sa11} computed radial slim disk profiles and included inner disk edge torques while \cite{sa09b} investigated effects of varying local dissipation physics on spectra through ray tracing. However, these studies did not self-consistently couple radiative transfer to disk structure. \cite{sdm13} later used BHSPEC models \citep{dav06} that are TLUSTY-based to calculate slim disk spectra but this work did not consider different dissipation physics. 

\

\ni Finally, we intend to expand this exploratory study to a more fine grained array of dissipation profiles, especially since recent simulations displayed found significant variations in their shapes and detailed height dependences. We will build a large library of disk annuli spectra and structure models with dissipation power-law index $\zeta$ as a parameter. These models, together with relativistic transfer efforts currently underway, would enable fitting observed photon spectra and QPO power spectra to constrain possible magnetic stresses at inner disk edge as well as spatial distribution of local dissipation.

\

\ni The authors acknowledge support from the Office of Undergraduate Research at University of San Diego. We also benefited from discussions with O. Blaes, J. Dexter and N. Storch. We also thank the referee for insightful questions and suggestions that greatly improved the clarity of our manuscript.


\begin{thebibliography}{}

\bibitem[Abramowicz et al.(1988)]{ab88} Abramowicz, M. A., Czerny B., Lasota J. P., \& Szuszkiewicz, E., 1988, ApJ, 332, 646

\bibitem[Agol(1997)]{agol97} Agol, E. 1997, Ph.D. thesis, Univ. California at Santa Barbara

\bibitem[Agol \& Krolik(2000)]{ak00} Agol, E., \& Krolik, J. H., 2000, ApJ, 528, 161

\bibitem[Balbus and Hawley(1991)]{bh91}Balbus, S. A., \& Hawley, J. F. 1991, ApJ, 376, 214

\bibitem[Balbus and Hawley(1998)]{bh98}Balbus, S. A., \& Hawley, J. F. 1998, Rev. Mod. Phys., 70, 1

\bibitem[Blaes et al.(2001)]{bla01}Blaes, O., Hubeny, I., Agol, E., \& Krolik J.H. 2001, ApJ, 563, 560

\bibitem[Blaes et al.(2006)]{bla06}Blaes, O., Hirose, S., Krolik, J. H., \& Davis. S. W. 2006, ApJ, 645, 1402

\bibitem[Blaes, Hirose \& Krolik(2007)]{bla07}Cunningham, C.T, 1975, ApJ, 202, 788

\bibitem[Blaes et al.(2011)]{bla11}Blaes, O., Krolik. J. H., Hirose, S., \& Shabaltas, N. 2011, ApJ, 733, 110

\bibitem[Belloni, Sanna \& Mendez(2012)]{bel12}Belloni, T. M., Sanna, A., \& Mendez, M. 2012, MNRAS 426, 1701

%\bibitem[Cunningham(1975)]{c75}Blaes, O., Krolik. J. H., Hirose, S., \& Shabaltas, N. 2011, ApJ, 733, 110

\bibitem[Davis et al.(2005)]{dav05}Davis, S.W., Blaes, O., Hubeny, I., \& Turner, N. J. 2005, ApJ, 621, 372

\bibitem[Davis et al.(2009)]{dav09}Davis, S.W., Blaes, O., Hirose, S., \& Krolik, J. H., 2009, ApJ, 703, 569.

\bibitem[Davis, Done \& Blaes(2006)]{dav062}Davis, S.W., Done C., \& Blaes, O., 2006, ApJ, 647, 525.

\bibitem[Davis \& Hubeny(2006)]{dav06}Davis, S.W., \& Hubeny, I., 2006, ApJ, 164, 530.

\bibitem[Davis \& El-Abd(2019)]{dav19}Davis, S.W., \& El-Abd, S., 2019, ApJ, 874, 23.

%\bibitem[Davis, Stone \& Jiang(2012)]{dav12}Davis, S. W., Stone, J. M., \& Jiang, Y.-F. 2012, ApJS, 199, 9

\bibitem[Dexter \& Blaes(2014)]{db14}Dexter, J., Blaes, O. 2014, MNRAS, 438, 3352

\bibitem[Dezen \& Flores(2018)]{df18}Dezen, T., \& Flores. B.,  2018, ApJ, 861, 18

\bibitem[Dezen, Egger \& Mwansa(2019)]{dem19}Dezen, T., \& Egger. N., \& Mwansa, L.,  2019, ApJ, 887, 162

%\bibitem[Done \& Kubota(2006)]{dk06}Done C., \& Kubota A., 2006, MNRAS, 371, 1216

\bibitem[Done, Gierlinski \& Kubota(2007)]{dgk07}Done C., Gierlinski, M., \& Kubota A., 2007, A\&ARv, 15, 1

\bibitem[Grove et al.(1998)]{gr98}Grove, J. E. et al., 1998, ApJ, 500, 899

%\bibitem[Hirose, Blaes \& Krolik(2009)]{hbk09}Hirose, S., Blaes, O.,  Krolik, J. H. 2009, ApJ, 704, 781

\bibitem[Hirose, Krolik \& Blaes(2009)]{hkb09}Hirose, S., Krolik, J. H., \& Blaes, O. 2009, ApJ, 691, 16

\bibitem[Hirose, Krolik \& Stone(2006)]{hir06}Hirose, S., Krolik, J. H., \& Stone, J. M. 2006, ApJ, 640, 901

\bibitem[Huang et al.(2023)]{huang23}Huang, J., Jiang, Y. F., Feng, H., Davis, S. W., Stone, J. M., \& Middleton, M. J.  2023, ApJ, 945, 57

\bibitem[Hubeny et al.(2000)]{h00}Hubeny, I., Agol, E., Blaes, O., Krolik, J. H.  2000, ApJ, 533, 710

%\bibitem[Hubeny et al.(2001)]{h01}Hubeny, I., Blaes, O., Krolik, J. H., \& Agol, E. 2001, ApJ, 559, 680

\bibitem[Hubeny \& Hubeny(1998)]{h98}Hubeny, I., \& Hubeny, V. 1998, ApJ, 505, 558

\bibitem[Hubeny \& Lanz(1995)]{hl95}Hubeny, I., \& Lanz, T. 1995, ApJ, 439, 875

\bibitem[Jiang, Stone \& Davis(2012)]{jia12}Jiang, Y. F., Stone, J. M., \& Davis, S. W. 2012, ApJS, 199, 14

\bibitem[Jiang, Davis \& Stone(2016)]{jia16}Jiang, Y. F., Davis, S. W., \&  Stone, J. M. 2016, ApJ, 827, 10

\bibitem[Ingram \& Motta(2019)]{ingram19}Ingram, R. A., \& Motta, S. E. 2019, New. Astron. Rev, 85, 101524

\bibitem[Jiang, Stone \& Davis(2013)]{jia13}Jiang, Y. F., Stone, J. M., \& Davis, S. W. 2013, ApJ, 778, 65

\bibitem[Jiang, Stone \& Davis(2014)]{jia14}Jiang, Y. F., Stone, J. M., \& Davis, S. W., 2014, ApJ, 796, 106

%\bibitem[Jiang, Stone \& Davis(2014)]{jia14}Jiang, Y. F., Stone, J. M., \& Davis, S. W., 2014, ApJS, 213, 7

%\bibitem[Kolehmainen, Done \& Trigo(2011)]{kol11}Kolehmainen, M., Done, C., \& Trigo, M. D., 2011, MNRAS, 416, 311

%\bibitem[Krolik(1999)]{kro99}Krolik, J. H., \textit{Active Galactic Nuclei}, Princeton University Press, Princeton, NJ, 1999.

%\bibitem[Krolik, Hawley, \& Hirose(2005)]{kro05}Krolik, J. H., Hawley, J. F.,
%\& Hirose, S. 2005, ApJ, 622, 1008

\bibitem[Krolik, Hirose, \& Blaes(2007)]{kro07}Krolik, J. H., Hirose, S., \& Blaes, O. 2007, ApJ, 664, 1045

%\bibitem[Kubota et al.(2010)]{ku10}Kubota A., Done, C., Davis, S.W., Dotani, T., Mizuno, T., \& Ueda, Y., 2010, ApJ, 714, 860.

%\bibitem[Kubota \& Done(2004)]{kd04}Kubota, A., \& Done, C. 2004, MNRAS, 353, 980

%\bibitem[Lightman \& Eardley(1974)]{le74}Lightman, A. P., \& Eardley, D. M. 1974, ApJ, 187, L1

\bibitem[Ling \& Wheaton(2005)]{lw05}Ling, J. C., \& Wheaton, W. A. 2005, ApJ, 622, 492

\bibitem[McClintock \& Remillard(2006)]{mr06}McClintock, J. E., \& Remillard, R. A. 2006, in Compact Stellar X-Ray Sources, eds. W. H. G Lewin \& M. van der Klis (New York: Cambridge University Press), 179

\bibitem[McClintock(2006)]{mc06}McClintock, J. E. et al. 2006, ApJ, 652, 518

\bibitem[Miller et al.(2001)]{miller01}Miller, J. M. et al. 2001, ApJ, 563, 928

\bibitem[Mills et al.(2024)]{mills24}Mills, B. S., Davis, S. W., Jiang, Y. F., \& Middleton, M. J. 2024, ApJ, 974, 166

\bibitem[Noble, Krolik \& Hawley(2009)]{noble09} Noble, S. C., Krolik, J. H., \& Hawley, J. F. 2009, ApJ, 692, 411

\bibitem[Noble, Krolik \& Hawley(2010)]{noble10} Noble, S. C., Krolik, J. H., \& Hawley, J. F. 2010, ApJ, 711, 959

\bibitem[Novikov \& Thorne(1973)]{nt73}Novikov, I. D., \& Thorne, K. S. 1973, in Black Holes, eds. C. De Witt \& B. De Witt (New York: Gordon \& Breach), 343

%\bibitem[Revnivtsev, Trudolyubov \& Borozdin(2000)]{rtb00}Revnivtsev, M. G., Trudolyubov, S. P., Borozdin, K. N. 2000, MNRAS, 312, 151

%\bibitem[Reynolds \& Miller(2009)]{rm09}Reynolds C. S., Miller M. C., 2009, ApJ, 692, 869

%\bibitem[Riffert \& Herold(1995)]{rh95}Riffert, H., \& Herold, H. 1995, ApJ, 450, 508

\bibitem[Torok \& Stuchlik(2005)]{ts05}Torok, G., \& Stuchlik, Z. 2005, A\& A, 437, 775

\bibitem[Sadowski(2009a)]{sa09a}Sadowski, A. 2009, ApJS, 183, 171

\bibitem[Sadowski et al.(2009b)]{sa09b}Sadowski, A. et al. 2009, A\&A, 502, 7

\bibitem[Sadowski et al.(2011)]{sa11}Sadowski, A. et al. 2011, A\&A, 527, 17

\bibitem[Schnittman, Krolik \& Noble(2013)]{sc13}Schnittman, J. D., Krolik, J. H., \& Noble, S. C., 2013, ApJ, 769, 156

\bibitem[Shakura \& Sunyaev(1973)]{ss73}Shakura, N. I., \& Sunyaev, R. A. 1973, A\&A 24, 337.

%\bibitem[Straub et al.(2011)]{s11}Straub O. et. al. 2011, A\&A 533, A67

\bibitem[Straub, Done \& Middleton(2013)]{sdm13}Straub, O., Done, C., Middleton, M.,  2013, A\&A 533, A67

%\bibitem[Shakura \& Sunyaev(1976)]{ss76}Shakura, N. I., \& Sunyaev, R. A. 1976, MNRAS, 175, 613

\bibitem[Tao \& Blaes(2013)]{tb13}Tao, T., \& Blaes, O., 2013, ApJ, 770, 55 

\bibitem[Turner(2004)]{tur04}Turner, N. J. 2004, ApJ, 605, L45

%\bibitem[Zdziarski et al.(2001)]{zd01}Zdziarski, A. A. et al. 2001, ApJ, 554, L45

\bibitem[Uttley \& McHardy(2001)]{u01}Uttley, P., \& McHardy, I. M., 2001, MNRAS, 323, L26

\bibitem[Zhang et al.(2025)]{zhang25}Zhang, Y. et al. 2025, in press

\bibitem[Zhu et al.(2012)]{zhu12}Zhu, Y. et al. 2012, MNRAS, 424, 2504

\end{thebibliography}
\end{document}